# Electrical detection and nucleation of a magnetic skyrmion in a magnetic tunnel junction observed via *operando* magnetic microscopy

J. Urrestarazu Larrañaga[1], Naveen Sisodia[1,‡], Van Tuong Pham[1,5,†], Ilaria Di Manici[1], Aurélien Masseboeuf[1], Kevin Garello[1], Florian Disdier[1], Bruno Fernandez[5], Sebastian Wintz[2], Markus Weigand[3], Mohamed Belmeguenai[4], Stefania Pizzini[5], Ricardo Sousa[1], Liliana Buda-Prejbeanu[1], Gilles Gaudin[1], Olivier Boulle[1,*]

[1] Univ. Grenoble Alpes, CNRS, CEA, Grenoble INP, SPINTEC, 38000 Grenoble, France
[2] Helmholtz-Zentrum Berlin für Materialien und Energie GmbH, D-14109 Berlin, Germany
[3] Max Planck Institute for Intelligent Systems, Heisenbergstraße 3, 70569 Stuttgart, Germany
[4] LSPM (CNRS-UPR 3407), Université Paris 13, Sorbonne Paris Cité, 99 avenue Jean-Baptiste Clément, Villetaneuse 93430, France
[5] Univ. Grenoble Alpes, CNRS, Institut Néel, 38042 Grenoble, France

## Abstract

Magnetic skyrmions are topological spin textures which are envisioned as nanometre scale information carriers in magnetic memory and logic devices. The recent demonstration of room temperature stabilization of skyrmions and their current induced manipulation in industry compatible ultrathin films were first steps towards the realisation of such devices. However, important challenges remain regarding the electrical detection and the low-power nucleation of skyrmions, which are required for the read and write operations. Here, we demonstrate, using *operando* magnetic microscopy experiments, the electrical detection of a single magnetic skyrmion in a magnetic tunnel junction (MTJ) and its nucleation and annihilation by gate voltage via voltage control of magnetic anisotropy. The nucleated skyrmion can be further manipulated by both gate voltage and external magnetic field, leading to tunable intermediate resistance states. Our results unambiguously demonstrate the readout and voltage controlled write operations in a single MTJ device, which is a major milestone for low power skyrmion based technologies.

## Introduction

Magnetic skyrmions have raised a large interest in recent years due to their rich physics and promising applications in the field of memory and logics[1–4]. They combine topological stability and small dimensions, down to the nanometre scale, which confer on them particle-like behaviour. Their ability to be moved by an electrical current, can be exploited to store and manipulate information at the nanometre scale[2]. For instance, skyrmions are foreseen to be used as information carriers in memory devices such as three-terminal MRAM and racetrack shift registers, where dense trains of skyrmions are manipulated in tracks using electrical current[1,2,5]. In addition to memory technology, skyrmions were a proposed as the basis of logic gates that leverage their repulsion to perform Boolean[6,7] or non-conventional computing operations[4], including stochastic[8], neuromorphic[9,10] or reservoir computing[11].

---

[*] E-mail : Olivier.boulle@cea.fr
[†] now at IMEC, 3001 Leuven, Belgium
[‡] now at Department of Physics, Indian Institute of Technology Gandhinagar, Gandhinagar-382355, Gujarat, India

Recent progress has brought us closer to the practical implementation of skyrmion-based devices, with the successful demonstration of room temperature skyrmions in ultrathin films[12–15] and their fast manipulation by an electrical current in tracks[15–19]. Despite these promising developments, fundamental challenges still remain concerning the electrical detection and controlled nucleation at low power, which are required for the read and write operations in devices. Pioneer works demonstrated the electrical detection of skyrmions at cryogenic temperature in B20 chiral magnets[20–24] using the topological Hall effect and in ultrathin films by spin polarized tunnel electron microscopy[25–27]. Later, single skyrmions were detected electrically at room temperature in ultrathin films using the anomalous Hall effect[28,29] or anomalous Nernst effect[30,31] but the low electrical signals (typically mΩ/skyrmion using Hall effect) prevent their implementation in real devices. A much larger readout signal is expected by exploiting the tunnelling magnetoresistance (TMR) effect in a magnetic tunnel junction (MTJ). Recent works reported distinct electrical signals in MTJs induced by magnetic field[32–34], gate voltage[35,36] or spin transfer torque[32], which were associated with the nucleation of magnetic skyrmions[32,33,35–37]. However, the lack of *operando* magnetic microscopy experiments did not allow for a direct, unambigous correspondence between the measured electrical and the underlying skyrmionic spin textures. More recently, the electrical detection of single skyrmions combined with its *in situ* observation was performed in MTJs using magnetic force microscopy (MFM)[38]. However, the use of a relatively thick skyrmion layer for the MFM observation prevented the electrical manipulation of the skyrmions, which is needed for any write operation.

In this work, we carefully optimized MTJ stack to stabilize magnetic skyrmions and achieve their electrical control using gate voltage. We employed x-ray magnetic microscopy experiments combined with *in-situ* electrical magneto-transport measurements to unambiguously demonstrate the electrical detection of a single magnetic skyrmion in an MTJ as well as its nucleation/annihilation using a gate voltage. The nucleation is driven by the voltage control of the magnetic anisotropy[39–41] (VCMA) in the MTJ. Furthermore, we show that the skyrmion can be manipulated using both external magnetic field and gate voltage. Our demonstration of the readout and write operations of a skyrmion in a MTJ is an important milestone towards the practical implementation of skyrmion based devices. It also offers promising pathways for the utilization of skyrmions in MTJ for low power non –conventional computing devices[9,11,42,43].

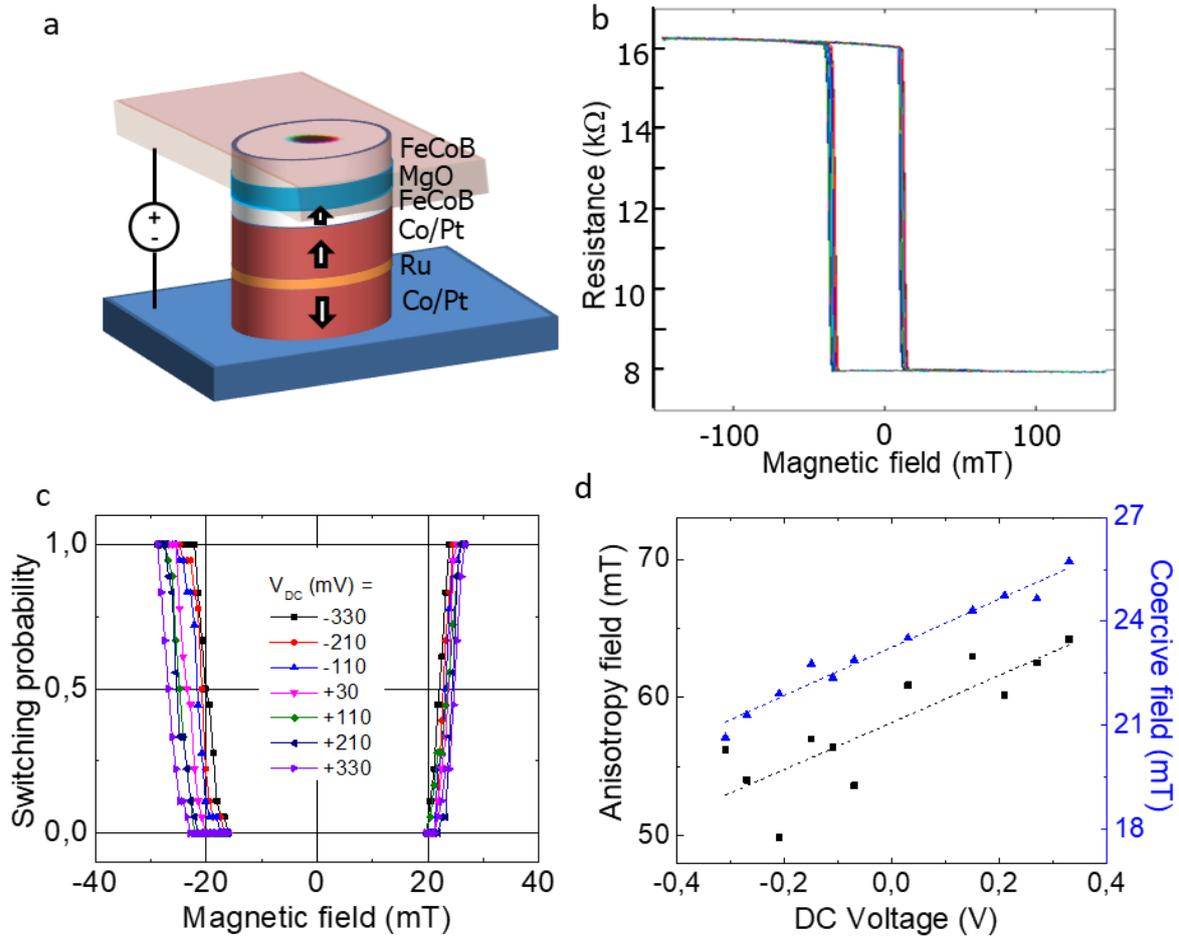

*Figure 1 **Magnetic tunnel junction for skyrmion stabilization and voltage control of magnetic anisotropy**. **a** Schematic representation of the MTJ pillar **b** Resistance vs out-of-plane magnetic field of a 150 nm diameter MTJ pillar. 18 hysteresis cycles are shown. A low DC voltage of 25 mV is applied. **c** Switching probability vs external magnetic field for different applied voltages. The switching from the P to AP configuration occurs for negative fields and from AP to P for negative fields. An offset field of 12 mT is subtracted. Each point is obtained by averaging 18 hysteresis loops. **d** Coercive field (blue triangle) and anisotropy field (black dots) vs DC voltage. The dotted lines are linear fits.*

# Magnetic tunnel junction for electrical skyrmion detection and nucleation

To stabilize a magnetic skyrmion and achieve their electrical detection and nucleation, the following MTJ stack (Figure 1a) was optimized: Ta/Pt/[Co/Pt]x6/Co/Ru/[Co/Pt]x3/Co/W/FeCoB/MgO/FeCoB/W/Pt. The stack is composed of a top FeCoB free layer and a bottom reference FeCoB layer coupled to a (Co/Pt) multilayer-based synthetic antiferromagnet to minimize the stray field. The stack was carefully optimized to nucleate the skyrmion with very low energy via VCMA[44]. To this end, we tuned the free layer thickness to achieve low perpendicular magnetic anisotropy (PMA), placing it close to the in-plane/out-of-plane spin reorientation transition (see ExtendedData Fig.1a). This optimization allows larger voltage induced change of the effective perpendicular magnetic anisotropy (PMA) while reducing the domain wall energy to favour the skyrmion nucleation. Furthermore, the thickness of the MgO barrier was adjusted to obtain a large resistance-area product of around 700 $\Omega.\mu m^2$. This enables us to control the free layer magnetic anisotropy via VCMA with minimal current and spin transfer torque, and hence to minimise the energy for skyrmion nucleation. Electrically contacted circular pillars were then fabricated using standard nanofabrication process, with diameters ranging between 50 nm and 2 μm. A TMR of around

100% is measured associated with a sharp reversal of the magnetisation when sweeping the magnetic field (Figure 1b, 150 nm diameter MTJ). The voltage dependence of the anisotropy field can be extracted from the switching probability vs magnetic field at different bias voltages[45,46] (see Figure 1c, Figure 1d). The linear dependence of the coercive and anisotropy fields on bias voltage (Figure 1d) demonstrates VCMA effect with a VCMA coefficient of 44 fJ/Vm, in line with the literature[39,46–48]. Complementary measurements of the anisotropy field at various voltages using hard axis hysteresis loops lead to similar values (Supplementary note 4). Dzyaoloshinskii-Moriya interaction (DMI) measurements using in-plane field skyrmion expansion as well as Brillouin Light Scattering experiments (BLS) reveal a low DMI of the free layer (D=-0.14±0.07 mJ/m², see Supplementary note 1). Micromagnetic simulations predict a Bloch domain wall structure but with a small Néel component (Supplementary note 5), a result in line with complementary Lorentz TEM experiments (Supplementary note 3).

## *Operando* STXM magnetic microscopy of MTJ

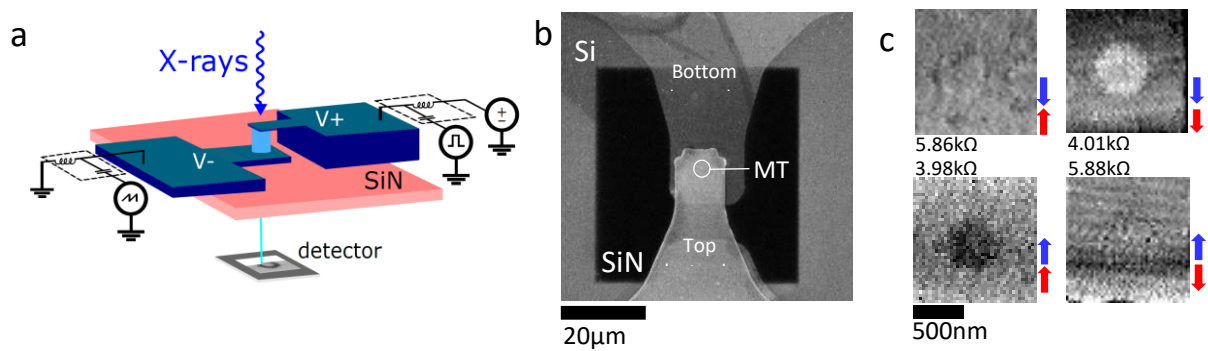

*Figure 2 **MTJ nanopillar for operando STXM magnetic microscopy. a** Schematic representation of the sample for STXM imaging. **b** Scanning electron microscopy (SEM) image of the sample. **c** XMCD-STXM images of the four possible single domain magnetisation configurations of the FeCoB layers in a 500 nm diameter MTJ (top left AP configurations measured at an out-of-plane magnetic field of 29.5 mT, top right P configuration measured at 250 mT, bottom left P configurations measured at -250 mT, bottom right AP configuration measured at -50 mT). The blue and red arrows stand for the magnetisation of the FeCoB free and reference layers respectively. Here the anti-parallel (AP) configurations appear as a grey contrast while the parallel (P) configurations appear as white or black contrast.*

To unambiguously demonstrate the skyrmion electrical detection and nucleation in the MTJ, it is essential to be able to simultaneously observe the free layer magnetisation and measure the MTJ resistance. This is a challenging task since it requires the high spatial resolution imaging of spin textures in an ultrathin film buried far from the sample surface. This cannot be achieved using standard table-top magnetic microscopy techniques such as magneto-optical Kerr effect, which can probe layers only a few nanometres from the surface for metallic stacks, or MFM, with a limited sensitivity and spatial resolution for nanometre thick films located far from the probe. In contrast, the relatively long penetration depth of soft x-rays in the materials commonly used in MTJ stacks makes scanning transmission x-ray microscopy (STXM) highly suited for *operando* magnetic imaging in MTJs[49,50]. To this end, the MTJ stack was grown on a 100 mm Si wafer where ultrathin (75 nm) $Si_3N_4$ membranes were locally patterned on 50x50 µm² windows. Sub-micron pillars were then fabricated on the membranes using standard MRAM nanofabrication processes (see Figure 2(a,b)). To allow for the x-ray transmission, only light elements, such as Ti and Al were used to contact the pillar electrically. For the STXM imaging, x-ray magnetic circular dichroism (XMCD) was used as magnetic contrast mechanism[51]. XMCD-STXM imaging combined with *in-situ* magneto-transport measurements confirms the observation of the FeCoB layer magnetisations at the Fe L3 edge and the simultaneous measurement of the MTJ resistance. As an example, we show in Figure 2c single domain magnetisation state images of the four possible configurations of the FeCoB magnetisation in a 500 nm diameter MTJ with their associated resistance state.

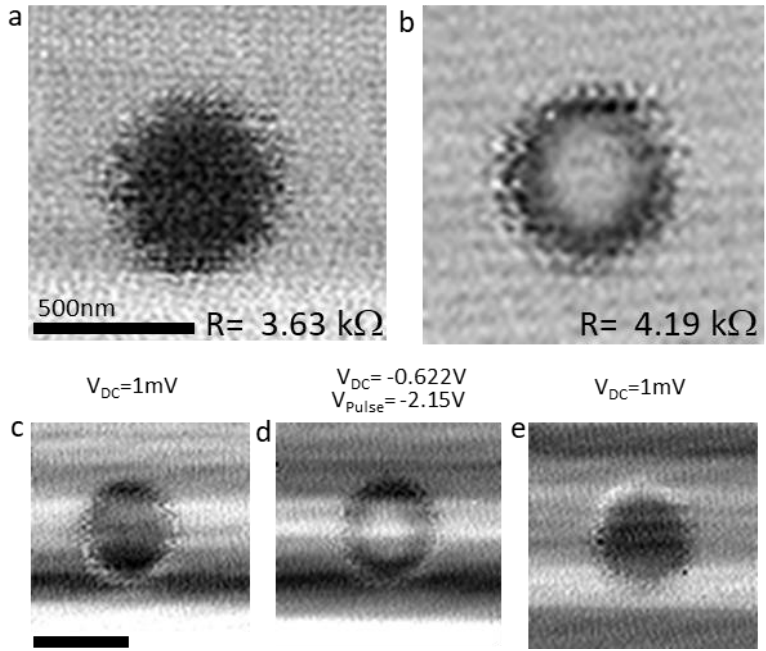

*Figure 3 **Electrical detection of a skyrmion in an MTJ and its nucleation/annihilation using gate voltage.** **a** XMCD-STXM image of a monodomain state measured at the Fe L3 edge in 500 nm diameter MTJ and its associated resistance state (parallel magnetisation configuration) (external perpendicular magnetic field $\mu_0H_z$=0 mT). **b** XMCD-STXM image measured after the application of a 10 ns 2.15 V voltage pulse applied from an initial parallel resistance state ($\mu_0H_z$=-2.75 mT). **c-e** STXM images **c** for a low DC voltage of 1mV, **d** after the application of a 10 ns voltage pulse of -2.15 V in the presence of a DC voltage of -622 mV and **e** after setting the DC voltage amplitude back to 1 mV. In c-e, $\mu_0H_z$=-2 mT. The black (grey) contrast shows that both FeCoB layers are in parallel (antiparallel) direction. The low 1mV DC voltage was used to measure the DC MTJ resistance. Black (grey) contrast areas on the image have their magnetisation aligned along positive (negative) magnetic field respectively. Negative (positive) voltage decreases (increases) the PMA respectively (cf Figure1). The scale bar in **c** is 500 nm.*

Only single domain states were observed when sweeping the magnetic field, in agreement with the measured square TMR hysteresis loops with sharp reversal (see Figure 3a of a XMCD-STXM image of a parallel state configuration at 0 mT in a 500 nm diameter MTJ and an example of a hysteresis loop in ExtendedData Fig1b). To nucleate magnetic skyrmions, we exploited the VCMA effect. Starting from an initial single domain state with parallel magnetisation configuration (R=3.73 kΩ, $\mu_0H_z$=-2.75 mT), the application of a 10 ns voltage pulse with an amplitude of -2.15V leads to the nucleation of a skyrmion in the FeCoB free layer (see Figure 3b). Here the negative voltage pulse leads to a decrease the PMA of the MTJ free layer (see Figure 1). The nucleation is associated with an increase of the resistance of 467 Ω, which is a quarter of the total TMR (ΔR=1.92 kΩ, TMR=51% in this sample). Note that this intermediate resistance is not observed in the hysteresis loop when sweeping the external magnetic field. This change in resistance is in agreement with that expected from the skyrmion diameter (240 nm) assuming parallel conduction channels for the P and AP resistance states (see Supplementary note 5.1). Due to the large RA product, a low current density, around $7\times10^9$ A/m², is injected when applying the voltage pulse. As a result, the spin transfer torque effect is expected to be negligible and the nucleation can be explained by the sole effect of the VMCA. A nucleation energy of around 30 pJ is estimated, which is large compared to the write energy in optimized STT-MRAM cells (~0.5 pJ)[52]. This difference is explained by the larger surface area of the MTJ dots, compared to the MTJ pillars of a few tens of nm in STT-MRAM. However, a much lower energy could be achieved by increasing further the RA product and/or reducing the MTJ size.

The obtained skyrmionic state is found to be stable against subsequent pulse voltage excitations, whether positive or negative. The range of magnetic fields in which the skyrmion remains stable is quite narrow (a few mT), the skyrmion diameter varying between 200 nm and 350 nm in this range (see Supplementary note 5.1). While the annihilation of the skyrmion was not observed using voltage pulses, a full electrical nucleation and annihilation was achieved by adding a small DC voltage to stabilize the skyrmion at a magnetic field value where it is otherwise unstable (see Figure 3(c-e), $\mu_0 H_z$=-2 mT). Starting from a single domain state (Figure 3c), a skyrmion is nucleated by applying a 10 ns voltage pulse in the presence of a DC voltage (Figure 3d), which further stabilizes the skyrmion. When releasing the DC voltage, the skyrmion state is destabilized and the initial monodomain state is again obtained (Figure 3e). This sequence demonstrates the full electrical nucleation/annihilation of a skyrmion in the MTJ.

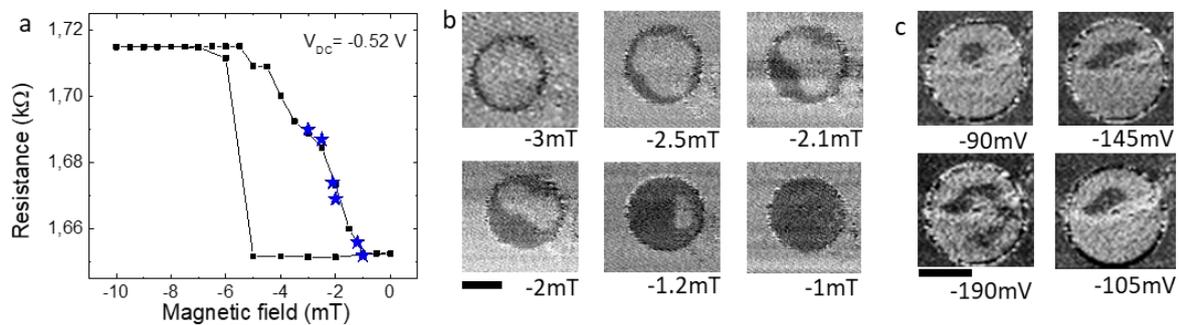

*Figure 4 **Magnetic field and bias voltage dependence of the skyrmion in a 2 µm MTJ pillar.** **a** Resistance vs perpendicular magnetic field hysteresis loop for a DC voltage of $V_{DC}$ = -0.52 V. The blue stars indicate the resistance values corresponding to the images shown in b **b** STXM images of the MTJ free layer (Fe L3 edge) for different magnetic field at $V_{DC}$ = -0.52 V. **c** Voltage dependence of the magnetic texture in the MTJ at a field of -2.3 mT. The scale bar in b and c is 1 µm. The images in b and c are normalized with respect to a STXM image of the pillar in the AP resistance state at a field of -20 mT. In a, an offset was applied to take into account a small drift in resistance during measurement compared to the black curve.*

In larger dots (2 µm diameter), the application of a negative DC voltage leads to a significant change of the hysteresis loop: While at low voltage a square hysteresis loop with sharp reversal is observed (not shown), a continuous decrease of the resistance from the AP to the P state is measured at larger voltage (Figure 4a, $V_{DC}$=-0.52V). STXM measurements reveal that this decrease in resistance is associated with the creation of a large skyrmion in the dot, with its domain wall close to the rim, which gradually decreases in size and eventually annihilates as the magnetic field increases toward positive values (*i.e* opposite to the skyrmion core magnetization) (Figure 4b). The nucleated skyrmion is also sensitive to the gate voltage amplitude (Figure 4c): an increase (decrease) of the skyrmion size is observed when increasing (decreasing) the voltage amplitude respectively, which is explained by the modulation of the domain wall energy via VCMA. The skyrmion shape can also be manipulated using nanosecond voltage pulses (see Supplementary note 5.2). The larger susceptibility of the skyrmion to external magnetic fields or gate voltages in larger dots can be explained by the shallower stray field energy potential when increasing the dot size[13,53]. Some events with unexpected bias dependence are also observed suggesting a partially stochastic response to the gate voltage. These events may be attributed to jumps enabled by voltage excitations exceeding energy barriers between the local energy minima defined by the dots geometry and the local pinning sites (see Supplementary note 5.2). The irregular shape of the skyrmion also evidences the influence of pinning[13,53].

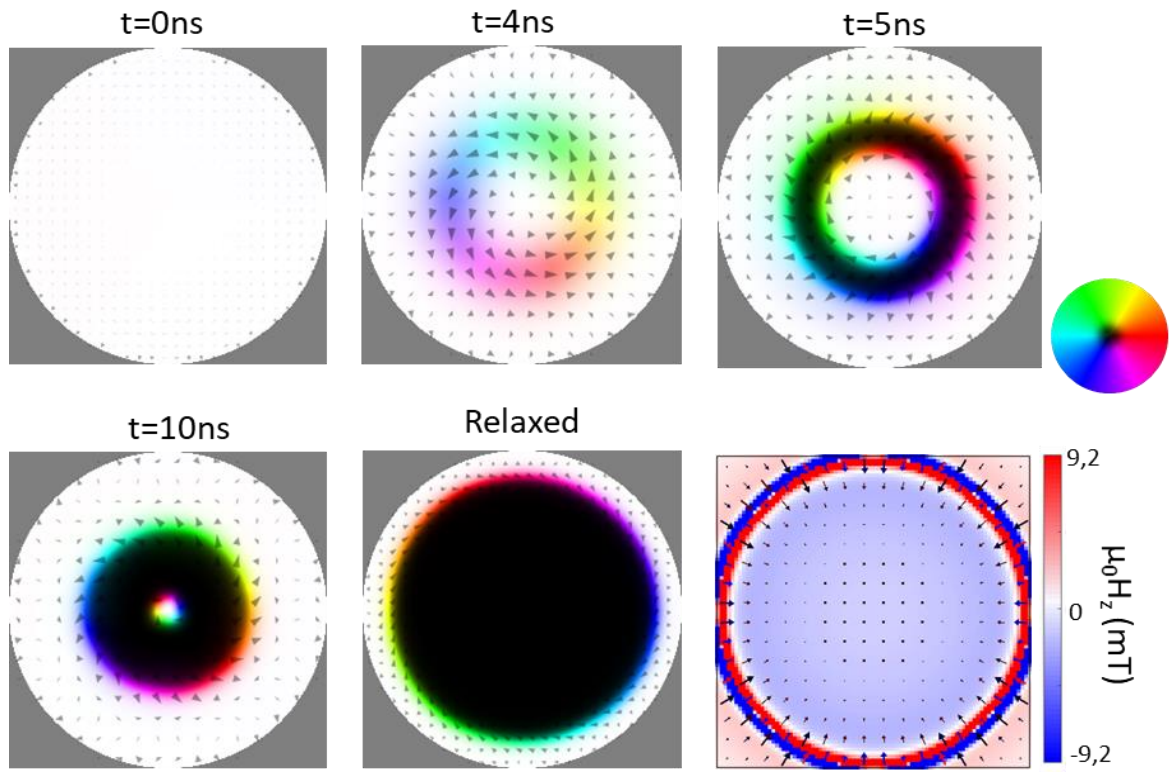

*Figure 5 **Micromagnetic simulation of the skyrmion nucleation.** Spatial distribution of the spin texture at different times in a 500 nm diameter dot during the application of a voltage pulse. The voltage pulse has an amplitude of -2.15 V and is applied at t=0 and switched off at 10 ns (400 ps rise time). Black/white contrast indicates the magnetisation pointing downward/upward. The colour wheel illustrates the orientation of the in-plane magnetisation. The bottom right panel shows the calculated stray field from the SAF layer, with the reference layer magnetisation being aligned parallel to the one of the free layer.*

## Micromagnetic simulations

To better understand these results, we performed micromagnetic simulations using experimental parameters. The applied voltage is assumed to decrease the perpendicular magnetic anisotropy via the VCMA. The spin transfer torque is neglected, due to the low injected current density (maximum $7 \times 10^9$ A/m²) resulting from the large R.A product. Figure 5 shows different snapshots of the magnetisation in a 500 nm dot during the application of a 10 ns voltage pulse of -2.15V. The nucleation proceeds via the creation of a transient curling state due to the decrease of the magnetic anisotropy (t=4 ns), followed by the nucleation of a ring with reversed perpendicular magnetisation (t=5 ns), a configuration which is favoured by the stray field energy. When the pulse is turned off (t=10 ns), the ring increases in size and leads to a single skyrmion, which eventually grows to its final size. As mentioned earlier, micromagnetic simulations predict a Bloch type skyrmion with a small Néel component owing to the low DMI in the stack. The simulations also show that the stray field from the SAF leads to a local field gradient close to the edge of the dot (see Figure 5, bottom right panel) which helps to stabilize the nucleated skyrmion and results in an enhancement of the field stability of the nucleated skyrmion (see Supplementary note 5.1). This local stray field can explain that experimentally the nucleation was only observed for a parallel magnetisation configuration (in 500 nm pillars), where the stray field stabilizes the skyrmion.

## Conclusion

To conclude, we have demonstrated the electrical detection and electrical nucleation and annihilation of a single skyrmion in a magnetic tunnel junction (MTJ) using *operando* magnetic microscopy experiments. The nucleation of the skyrmion was enabled by voltage excitations exploiting the VCMA effect and leads to a large variation of the MTJ resistance (470 Ω) owing to the large TMR of the junction. Micromagnetic simulations using experimental parameters reproduce the nucleation process well and show that the nucleation proceeds via the transient creation of a curling state. The skyrmion can be manipulated using voltage excitations and the external magnetic field, which leads to tunable intermediate resistance states. Our results mark major milestones for skyrmion based devices, demonstrating unambiguously the readout and low power write operations in a single device. Furthermore, they open promising pathways for low power multilevel memory and neuromorphic skyrmionic devices, leveraging the tunable manipulation of the skyrmion state hosted in the MTJ using low power voltage excitation.

## Acknowledgement
We acknowledge the support of the French Agence Nationale de la Recherche, project ANR-17-CE24-0045 (SKYLOGIC), the DARPA TEE program through grant MIPR# HR0011831554 from the DOI. We would like to thank Martien den Hertog from Institut Néel, Grenoble, France for providing samples for the membrane preparation, the Helmholtz-Zentrum Berlin for the allocation of synchrotron radiation beamtime.

# Materials and Methods

## Sample preparation

The MTJ stacks were deposited by magnetron sputtering. The stacks had the following composition: Substrate/Ta(3)/Pt($t_{Pt\_seed}$)/[Co(0.5)/Pt(0.3)]x6/Co(0.5)/Ru(0.9)/[Co(0.5)/Pt(0.3)]x3/ Co(0.5)/W(0.2)/Fe$_{72}$Co$_8$B$_{20}$(1)/MgO/Fe$_{72}$Co$_8$B$_{20}$($t_{FCB}$)/W(1.5)/Pt(2), thicknesses in nm. The MgO layer was obtained by the natural oxidation of a Mg layer under an oxygen pressure of 150 mbar for 10 s such that MgO=Mg(0.9)/oxidation/Mg(0.9)/oxidation/Mg(0.5). The free Fe$_{72}$Co$_8$B$_{20}$ layer was deposited as a wedge such that its nominal thickness $t_{FCB}$ varies linearly along the 100mm wafer between 1.51 and 1.65 nm. Two samples were prepared. A first sample A for the TMR and VCMA characterization. This stack was deposited on a highly resistive Si wafer and the Pt seed layer had a thickness of $t_{Pt\_seed}$=10 nm. A second sample B for STXM measurements was prepared where the stack was deposited on a 300 μm thick Si/SiO$_2$(200 nm)/Si$_3$N$_4$(75 nm) wafer with a thinner Pt seed layer ($t_{Pt\_seed}$=5 nm) to limit X-ray absorption at the Fe L3 edge. After deposition, the samples were annealed at 300°C for 10 minutes under high vacuum.

ExtendedData Fig.1a shows the out-of-plane hysteresis loops measured by magneto-optical Kerr effect (MOKE) of a sample deposited on a Si substrate for different thicknesses of the FeCoB free layer ranging between 1.57 nm-1.61 nm, which corresponds to the region of interest for the STXM observation. The hysteresis loops display the typical bended and butterfly hysteresis loops associated with stripe domain configurations, which are expected close to the in-plane to out-of-plane reorientation transition. The nominal thickness of the FeCoB free layer of the sample shown in the main text was t=1.59nm.

The stack was then patterned into sub-micronic pillar using standard MRAM nanofabrication processes. In sample B, for the STXM observation, 50x50 μm$^2$ Si$_3$N$_4$ windows were defined before the pillar patterning using photolithography followed by KOH chemical etching. Special care was taken when designing the window pattern and during the nanofabrication to limit the stress on the Si$_3$N$_4$ membranes. Also, materials with little absorption at the L3 Co and Fe edges were used to contact the pillar, such as a 120 nm Ti layer used to define the pillar by ion beam etching and a 225 nm thick Al top metallic electrode.

ExtendedData Fig.1b shows the resistance vs perpendicular magnetic field hysteresis loop of a 500 nm diameter MTJ pillar fabricated on $Si_3N_4$ membranes and shown in Figure 2 of the main text. TMR > 53% with sharp reversal and a resistance area product around 700 Ω.µm² are measured. The lower TMR ratio compared to samples fabricated on Si wafer is due to the thinner Pt seed layer (5 nm) used to limit the X-ray absorption. As a result, the bottom electrode after the nanofabrication process is more resistive (around 1.22 kΩ) and contributes more to the total MTJ resistance. After correcting for the bottom electrode resistance, a TMR of 75 % is found.

## Characterization of the magnetic properties

To characterize the magnetic properties of the free FeCoB layer, a stack similar to the MTJ was deposited, in which only the free FeCoB layer was magnetically active. This was achieved by substituting the Co in the SAF with Pt and replacing the fixed FeCoB thickness with a magnetically dead FeCoB layer of 0.3 nm. This sample was used to characterize the magnetic parameters of the FeCoB free layer, namely, the magnetic moment, magnetic anisotropy, Dzyaloshinskii-Moriya interaction (DMI), magnetic damping. The magnetic moment and magnetic anisotropy were measured using SQUID magnetometry, the Dzyaloshinskii-Moriya interaction (DMI) using Brillouin light scattering experiments as well as magnetic bubble expansion, the magnetic damping and gyromagnetic ratio by ferromagnetic resonance experiments. All measurements are performed at room temperature. These results are described in ExtendedData Table1. The details of the experiments are described in supplementary Note 1.

## Voltage control magnetic anisotropy experiments

The voltage control of magnetic anisotropy (VCMA) effect in the MTJ was characterized using two different experiments. Firstly, from the voltage dependence of the out-of-plane magnetic field ($H_z$) switching probability (see Figure 1c of the main text) [45,46]. To this end, the probability curves were measured for different voltages, and were fitted with the following distribution[45,46]:

$P_{SW}(H_Z) = 1 - \exp\left(-\frac{H_{Keff}f_0\sqrt{\pi}}{2R\sqrt{\Delta}}\text{erfc}\left[\sqrt{\Delta}\left(1 - \frac{H_Z - H_f}{H_{Keff}}\right)\right]\right)$ where $f_0$ is the attempt frequency, $R$ is the external magnetic field sweep rate and $H_f$ is the stray field from the fixed FeCoB and SAF layers. This allows us to extract the magnetic anisotropy field $H_{Keff}$ and the thermal stability factor $\Delta$ as a function of the voltage (see Figure 1d of the main paper). A linear fit leads to a variation of the anisotropy of $\frac{dH_{K,eff}}{dV} = 17$ mT/V. The VCMA coefficient $\xi$ writes $\xi = \frac{\mu_0 M_{S,FL} t_{FL} t_{MgO}}{2}\frac{dH_{K,eff}}{dV}$ where $M_{S,FL}$ refers to the saturation magnetisation of the free layer, $t_{FL}$ is its thickness and $t_{MgO}$ is the barrier thickness. Here we use $t_{MgO}$=3.03 nm, $t_{FL}$=1.04 nm and $M_{S,FL}$=1.63 MA/m leading to a value of $\xi$ =44 fJ/V. We describe in the supplementary Note 4 another method based on hard axis magneto-transport measurements which leads to a similar estimation of the VCMA coefficient.

## Micromagnetic simulations

The micromagnetic simulations were carried out using Mumax3[54,55]. The mesh size was 1.95 nm. The following micromagnetic parameters were used: saturation magnetisation $M_S$ =1.63 MA/m, perpendicular anisotropy field $\mu_0 H_K$=30 mT, DMI parameter D=-0.148 mJ/m², VCMA coefficient=16.2 mT/V, magnetic damping $\alpha$=0.016. These parameters were derived experimentally (see Supplementary note 1). An exchange interaction parameters A=12 pJ/m was used[56]. The simulations include the stray field generated by the FeCoB reference layer and by the Pt/Co SAF. It was simulated independently using experimental parameters, assuming the different layers composing the SAF have opposite magnetisation aligned along z.

## Scanning transmission X-ray microscopy

Scanning transmission X-ray magnetic microscopy experiments were carried out at the Maxymus endstation at the Bessy II electron storage ring operated by the Helmholtz-Zentrum Berlin für Materialien und Energie. All experiments were performed at room temperature. The X-ray beam was impinging perpendicularly to the sample surface, such that the logarithmic transmission magnetic contrast is proportional to the z component of magnetisation, normal to the sample plane. The voltage induced skyrmion nucleation was found to be highly reproducible in some range of magnetic field and DC voltages and was observed in four different samples (two 500 nm diameter and two 2µm diameter MTJs).

## Reference

1. Fert, A., Cros, V. & Sampaio, J. Skyrmions on the track. *Nat Nano* **8**, 152–156 (2013).

2. Fert, A., Reyren, N. & Cros, V. Advances in the Physics of Magnetic Skyrmions and Perspective for Technology. *Nature Reviews Materials* **2**, 17031 (2017).

3. Nagaosa, N. & Tokura, Y. Topological properties and dynamics of magnetic skyrmions. *Nat Nano* **8**, 899–911 (2013).

4. Vakili, H. *et al.* Skyrmionics—Computing and memory technologies based on topological excitations in magnets. *Journal of Applied Physics* **130**, 070908 (2021).

5. Cros, V., Fert, A., SAMPAIO, J. & SENEOR, P. Dispositif memoire avec skyrmions magnetiques et procede associe. (2022).

6. Sisodia, N. *et al.* Robust and Programmable Logic-In-Memory Devices Exploiting Skyrmion Confinement and Channeling Using Local Energy Barriers. *Phys. Rev. Applied* **18**, 014025 (2022).

7. Sisodia, N. *et al.* Programmable Skyrmion Logic Gates Based on Skyrmion Tunneling. *Phys. Rev. Applied* **17**, 064035 (2022).

8. Pinna, D. *et al.* Skyrmion Gas Manipulation for Probabilistic Computing. *Phys. Rev. Applied* **9**, 064018 (2018).

9. Song, K. M. *et al.* Skyrmion-based artificial synapses for neuromorphic computing. *Nature Electronics* **3**, 148–155 (2020).

10. Huang, Y., Kang, W., Zhang, X., Zhou, Y. & Zhao, W. Magnetic skyrmion-based synaptic devices. *Nanotechnology* **28**, 08LT02 (2017).

11. Pinna, D., Bourianoff, G. & Everschor-Sitte, K. Reservoir Computing with Random Skyrmion Textures. *Phys. Rev. Applied* **14**, 054020 (2020).

12. Jiang, W. *et al.* Blowing magnetic skyrmion bubbles. *Science* **349**, 283–286 (2015).

13. Boulle, O. *et al.* Room-temperature chiral magnetic skyrmions in ultrathin magnetic nanostructures. *Nat Nano* **11**, 449–454 (2016).

14. Moreau-Luchaire, C. *et al.* Additive interfacial chiral interaction in multilayers for stabilization of small individual skyrmions at room temperature. *Nat Nano* **11**, 444–448 (2016).

15. Woo, S. *et al.* Observation of room-temperature magnetic skyrmions and their current-driven dynamics in ultrathin metallic ferromagnets. *Nat Mater* **15**, 501–506 (2016).

16. Juge, R. *et al.* Current-Driven Skyrmion Dynamics and Drive-Dependent Skyrmion Hall Effect in an Ultrathin Film. *Phys. Rev. Applied* **12**, 044007 (2019).

17. Jiang, W. *et al.* Direct observation of the skyrmion Hall effect. *Nature Phys* **13**, 162–169 (2017).

18. Litzius, K. *et al.* Skyrmion Hall effect revealed by direct time-resolved X-ray microscopy. *Nat Phys* **13**, 170–175 (2017).

19. Litzius, K. *et al.* The role of temperature and drive current in skyrmion dynamics. *Nat Electron* **3**, 30–36 (2020).

20. Neubauer, A. *et al.* Topological Hall Effect in the A Phase of MnSi. *Phys. Rev. Lett.* **102**, 186602 (2009).

21. Yin, G., Liu, Y., Barlas, Y., Zang, J. & Lake, R. K. Topological spin Hall effect resulting from magnetic skyrmions. *Phys. Rev. B* **92**, 024411 (2015).

22. Porter, N. A., Gartside, J. C. & Marrows, C. H. Scattering mechanisms in textured FeGe thin films: Magnetoresistance and the anomalous Hall effect. *Phys. Rev. B* **90**, 024403 (2014).

23. Schulz, T. *et al.* Emergent electrodynamics of skyrmions in a chiral magnet. *Nature Phys* **8**, 301–304 (2012).

24. Kanazawa, N. *et al.* Large Topological Hall Effect in a Short-Period Helimagnet MnGe. *Phys. Rev. Lett.* **106**, 156603 (2011).

25. Romming, N. *et al.* Writing and Deleting Single Magnetic Skyrmions. *Science* **341**, 636–639 (2013).

26. Hanneken, C. *et al.* Electrical detection of magnetic skyrmions by tunnelling non-collinear magnetoresistance. *Nature Nanotechnology* **10**, 1039–1042 (2015).

27. Perini, M. *et al.* Electrical Detection of Domain Walls and Skyrmions in Co Films Using Noncollinear Magnetoresistance. *Phys. Rev. Lett.* **123**, 237205 (2019).

28. Maccariello, D. *et al.* Electrical detection of single magnetic skyrmions in metallic multilayers at room temperature. *Nature Nanotechnology* **13**, 233 (2018).

29. Zeissler, K. *et al.* Discrete Hall resistivity contribution from Néel skyrmions in multilayer nanodiscs. *Nature Nanotech* **13**, 1161–1166 (2018).

30. Wang, Z. *et al.* Thermal generation, manipulation and thermoelectric detection of skyrmions. *Nat Electron* **3**, 672–679 (2020).

31. Fernández Scarioni, A. *et al.* Thermoelectric Signature of Individual Skyrmions. *Phys. Rev. Lett.* **126**, 077202 (2021).

32. Penthorn, N. E., Hao, X., Wang, Z., Huai, Y. & Jiang, H. W. Experimental Observation of Single Skyrmion Signatures in a Magnetic Tunnel Junction. *Phys. Rev. Lett.* **122**, 257201 (2019).

33. He, B. *et al.* Realization of Zero-Field Skyrmions in a Magnetic Tunnel Junction. *Advanced Electronic Materials* **n/a**, 2201240.

34. Li, S. *et al.* Experimental demonstration of skyrmionic magnetic tunnel junction at room temperature. *Science Bulletin* (2022) doi:10.1016/j.scib.2022.01.016.

35. Kasai, S., Sugimoto, S., Nakatani, Y., Ishikawa, R. & Takahashi, Y. K. Voltage-controlled magnetic skyrmions in magnetic tunnel junctions. *Appl. Phys. Express* **12**, 083001 (2019).

36. Chen, S. *et al.* All-Electrical Skyrmionic Bits in a Chiral Magnetic Tunnel Junction. Preprint at https://doi.org/10.48550/arXiv.2302.08020 (2023).

37. Li, S. *et al.* Experimental demonstration of skyrmionic magnetic tunnel junction at room temperature. *Science Bulletin* (2022) doi:10.1016/j.scib.2022.01.016.

38. Guang, Y. *et al.* Electrical Detection of Magnetic Skyrmions in a Magnetic Tunnel Junction. *Advanced Electronic Materials* **n/a**, 2200570.

39. Khalili Amiri, P. *et al.* Electric-Field-Controlled Magnetoelectric RAM: Progress, Challenges, and Scaling. *IEEE Transactions on Magnetics* **51**, 1–7 (2015).

40. Nozaki, T. *et al.* Recent Progress in the Voltage-Controlled Magnetic Anisotropy Effect and the Challenges Faced in Developing Voltage-Torque MRAM. *Micromachines (Basel)* **10**, 327 (2019).

41. Kanai, S. *et al.* Electric field-induced magnetization reversal in a perpendicular-anisotropy CoFeB-MgO magnetic tunnel junction. *Applied Physics Letters* **101**, 122403 (2012).

42. Liang, X. *et al.* A spiking neuron constructed by the skyrmion-based spin torque nano-oscillator. *Appl. Phys. Lett.* **116**, 122402 (2020).

43. Taniguchi, T., Ogihara, A., Utsumi, Y. & Tsunegi, S. Spintronic reservoir computing without driving current or magnetic field. *Sci Rep* **12**, 10627 (2022).

44. Schott, M. *et al.* The Skyrmion Switch: Turning Magnetic Skyrmion Bubbles on and off with an Electric Field. *Nano Lett.* **17**, 3006–3012 (2017).

45. Thomas, L. *et al.* Perpendicular spin transfer torque magnetic random access memories with high spin torque efficiency and thermal stability for embedded applications (invited). *Journal of Applied Physics* **115**, 172615 (2014).

46. Wu, Y. C. *et al.* Voltage-Gate-Assisted Spin-Orbit-Torque Magnetic Random-Access Memory for High-Density and Low-Power Embedded Applications. *Phys. Rev. Applied* **15**, 064015 (2021).

47. Nozaki, T. *et al.* Recent Progress in the Voltage-Controlled Magnetic Anisotropy Effect and the Challenges Faced in Developing Voltage-Torque MRAM. *Micromachines* **10**, 327 (2019).

48. Dieny, B. & Chshiev, M. Perpendicular magnetic anisotropy at transition metal/oxide interfaces and applications. *Rev. Mod. Phys.* **89**, 025008 (2017).

49. Yu, X. W. *et al.* Images of a Spin-Torque-Driven Magnetic Nano-Oscillator. *Phys. Rev. Lett.* **106**, 167202 (2011).

50. Bernstein, D. P. *et al.* Nonuniform switching of the perpendicular magnetization in a spin-torque-driven magnetic nanopillar. *Phys. Rev. B* **83**, 180410 (2011).

51. Schütz, G. *et al.* Absorption of circularly polarized x rays in iron. *Phys. Rev. Lett.* **58**, 737–740 (1987).

52. Dieny, B. *et al.* Opportunities and challenges for spintronics in the microelectronics industry. *Nature Electronics* **3**, 446–459 (2020).

53. Juge, R. *et al.* Magnetic skyrmions in confined geometries: Effect of the magnetic field and the disorder. *Journal of Magnetism and Magnetic Materials* **455**, 3–8 (2018).

54. Vansteenkiste, A. *et al.* The design and verification of MuMax3. *AIP Advances* **4**, 107133 (2014).

55. Vansteenkiste, A. & Van de Wiele, B. MuMax: A new high-performance micromagnetic simulation tool. *Journal of Magnetism and Magnetic Materials* **323**, 2585–2591 (2011).

56. Srivastava, T. *et al.* Large-Voltage Tuning of Dzyaloshinskii–Moriya Interactions: A Route toward Dynamic Control of Skyrmion Chirality. *Nano Lett.* **18**, 4871–4877 (2018).

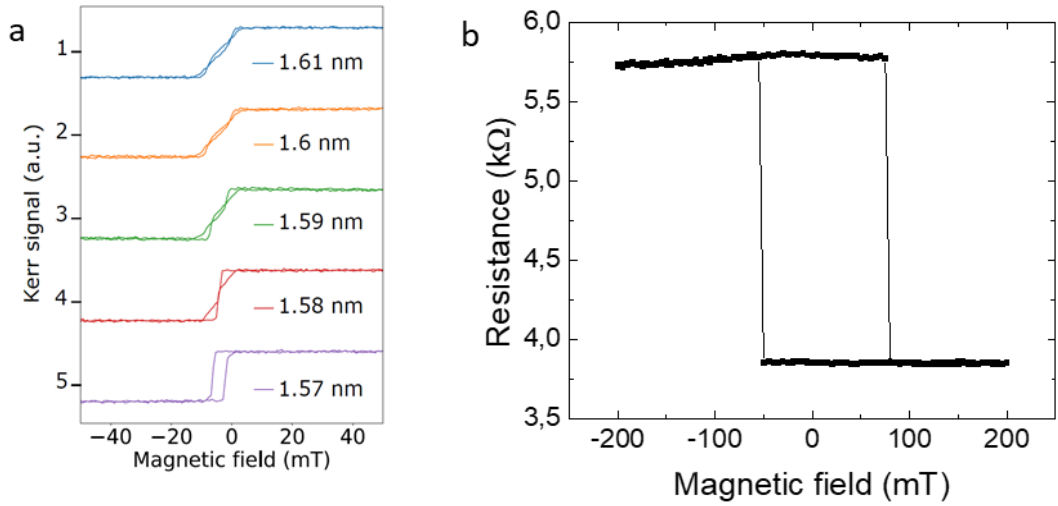

*ExtendedData Fig. 1* **Optimisation of the MTJ.** **a** Magneto-optical Kerr effect (MOKE) hysteresis loops of the FeCoB free layer for different thicknesses of the FeCoB layer. The magnetic field is applied perpendicularly to the sample plane. The curves are offset for clarity. **b** Resistance vs out-of-plane magnetic field hysteresis loop of a 500 nm diameter MTJ device fabricated on $Si_3N_4$ membranes. XMCD-STXM images of this device are shown in Figure 2 of the main text.

| | |
|---|---|
| $t_{FeCoB}$ (nm) | 1.59 |
| $t_{dead}$ (nm) | 0.51 |
| $M_S$ (MA/m) | 1.63±0.19 |
| $\mu_0 H_K$ (mT) | 35±5 |
| α (FMR) | 0.016±0.001 |
| D (mJ/m$^2$) (BLS) | -0.14±0.07 |
| γ/2π (GHz/T) | 29.6 |

*ExtendedData Table 1:* **Summary of the experimental parameters.** Nominal and dead layer FeCoB thicknesses, saturation magnetisation Ms, anisotropy field $\mu_0 H_K$, magnetic damping parameter, DMI constant and gyromagnetic ratio.